\documentclass[aps,prx,preprint,groupedaddress,amssymb,amsmath,floatfix]{revtex4-1} 

\usepackage{graphicx}
\usepackage{bm}
\usepackage{cleveref}

\usepackage{subfig}
\usepackage{color}

\begin{document}
\title{Photonic-Crystal Exciton-Polaritons in Monolayer Semiconductors}
\author{Long Zhang$^1$, Rahul Gogna$^2$, Will Burg$^3$, Emanuel Tutuc$^3$}
\author{Hui~Deng$^{1,2}$} \email{dengh@umich.edu}
\small{ }
\address{$^1$ Physics Department, University of Michigan, 450 Church Street, Ann Arbor, MI 48109-2122, USA}
\address{$^2$ Applied Physics Program, University of Michigan, 450 Church Street, Ann Arbor, MI 48109-1040, USA}
\address{$^3$ Microelectronics Research Center, Department of Electrical and Computer Engineering, The University of Texas at Austin, Austin, Texas 78758, United States
}


\begin{abstract}
Semiconductor microcavity polaritons, formed via strong exciton-photon coupling, provide a quantum many-body system on a chip, featuring rich physics phenomena for better photonic technology.
However, conventional polariton cavities are bulky, difficult to integrate, and inflexible for mode control, especially for room temperature materials. 
Here we demonstrate sub-wavelength thick one-dimensional photonic crystals (PCs) as a designable, compact and practical platform for strong coupling with atomically thin van der Waals Crystals (vdWCs). Polariton dispersions and mode anti-crossings are measured up to room temperature. Non-radiative decay to dark excitons was suppressed due to polariton enhancement of the radiative decay. 
Unusual features, including highly anisotropic dispersions and adjustable Fano resonances in reflectance, may facilitate high temperature polariton condensation in variable dimensions.
Combining slab PCs and vdWCs in the strong coupling regime allows unprecedented engineering flexibility for exploring novel polariton phenomena and device concepts.
\end{abstract}

\pacs{Valid PACS appear here}

\maketitle
\section{Introduction}
Control of light-matter interactions is elementary to the development of photonic devices. Existing photonic technologies are based on weakly coupled matter-light  systems, where the optical structure perturbatively modifies the electronic properties of the active media.
As the matter-light interaction becomes stronger and no longer perturbative, light and matter couple to form hybrid quasi-particles -- polaritons.
In particular, quantum-well (QW) microcavity exciton polaritons feature simultaneously strong excitonic nonlinearity, robust photon-like coherence, and a meta-stable ground state, providing a fertile ground for quantum many-body physics phenomena \cite{deng_exciton-polariton_2010,carusotto_quantum_2013} that promise new photonic technology \cite{fraser_physics_2016}.
%
Numerous novel types of many-body quantum states with polaritons and polariton quantum technologies have been conceived, such as topological polaritons \cite{karzig_topological_2015,bardyn_topological_2015,nalitov_spin-orbit_2015}, polariton neurons \cite{liew_optical_2008}, non-classical state generators \cite{liew_single_2010,flayac_single_2016,joana_steady-state_2016}, and quantum simulators \cite{byrnes_mott_2010,douglas_quantum_2015,angelakis_quantum_2017}. Their implementation require confined and coupled polariton systems with engineered properties, which, on one hand can be created by engineering the optical component of the strongly coupled modes, on the other hand, is difficult experimentally using conventional polariton systems.

Conventional polariton system are based on vertical FP cavities made of thick stacks of planar, distributed Bragg reflectors (DBRs), which have no free design parameter for mode-engineering and are relatively rigid and bulky against post-processing. Different cavity structures have been challenging to implement for polariton systems as conventional materials are sensitive to free surfaces and lattice mismatch with embedding crystals.
The recently emerged two-dimensional (2D) semiconductor vdWCs \cite{mak_atomically_2010,splendiani_emerging_2010} are uniquely compatible with diverse substrate without lattice matching \cite{geim_van_2013}. 
However, most studies of vdWC-polaritons so far continue to use FP cavities \cite{liu_strong_2014, dufferwiel_excitonpolaritons_2015,sidler_fermi_2016, flatten_room-temperature_2016,liu_control_2017,chen_valley-polarized_2017,dufferwiel_valley-addressable_2017}, which are even more limiting for vdWCs than for conventional materials. This is because  monolayer-thick vdWCs need to be sandwiched in between separately fabricated DBR stacks and positioned very close to the cavity-field maximum. The process is complex, hard to control, and may change or degrade the optical properties of vdWCs \cite{sercombe_optical_2013,chen_valley-polarized_2017}.
Alternatively, metal mirrors and plasmonic structures have been implemented \cite{wang_coherent_2016,liu_strong_2016,lundt_room-temperature_2016,sun_optical_2017}. They are more compact and flexible, but suffer from intrinsically large absorption loss and poor dipole-overlap between the exciton and field \cite{bhanu_photoluminescence_2015,wang_coherent_2016}.

Here we demonstrate sub-wavelength thick, one-dimensional dielectric PCs as a readily designable platform for strong-coupling, which is also ultra-compact, practical, and especially well suited to the atomically-thin vdWCs. Pristine vdWCs can be directly laid on top of the PC without further processing. Properties of the optical modes, and in turn the polariton modes, can be modified with different designs of the PC.
We confirm polariton modes up to room temperature by measuring the polariton dispersions and mode anti-crossing in both reflectance and photoluminescence (PL) spectra. Strongly suppressed non-radiative decay to dark excitons due to the polaritonic enhancement was observed. We show that these polaritons have anisotropic polariton dispersions and adjustable reflectance, suggesting greater flexibility in controlling the excitations in the system to reaching vdWC-polariton condensation at lower densities in variable dimensions.
Extension to more elaborate PC designs and 2D PCs will facilitate research on polariton physics and devices beyond 2D condensates.
\section{Results}
\subsection{The system}
\begin{figure*}[t]
	\includegraphics[width=\linewidth]{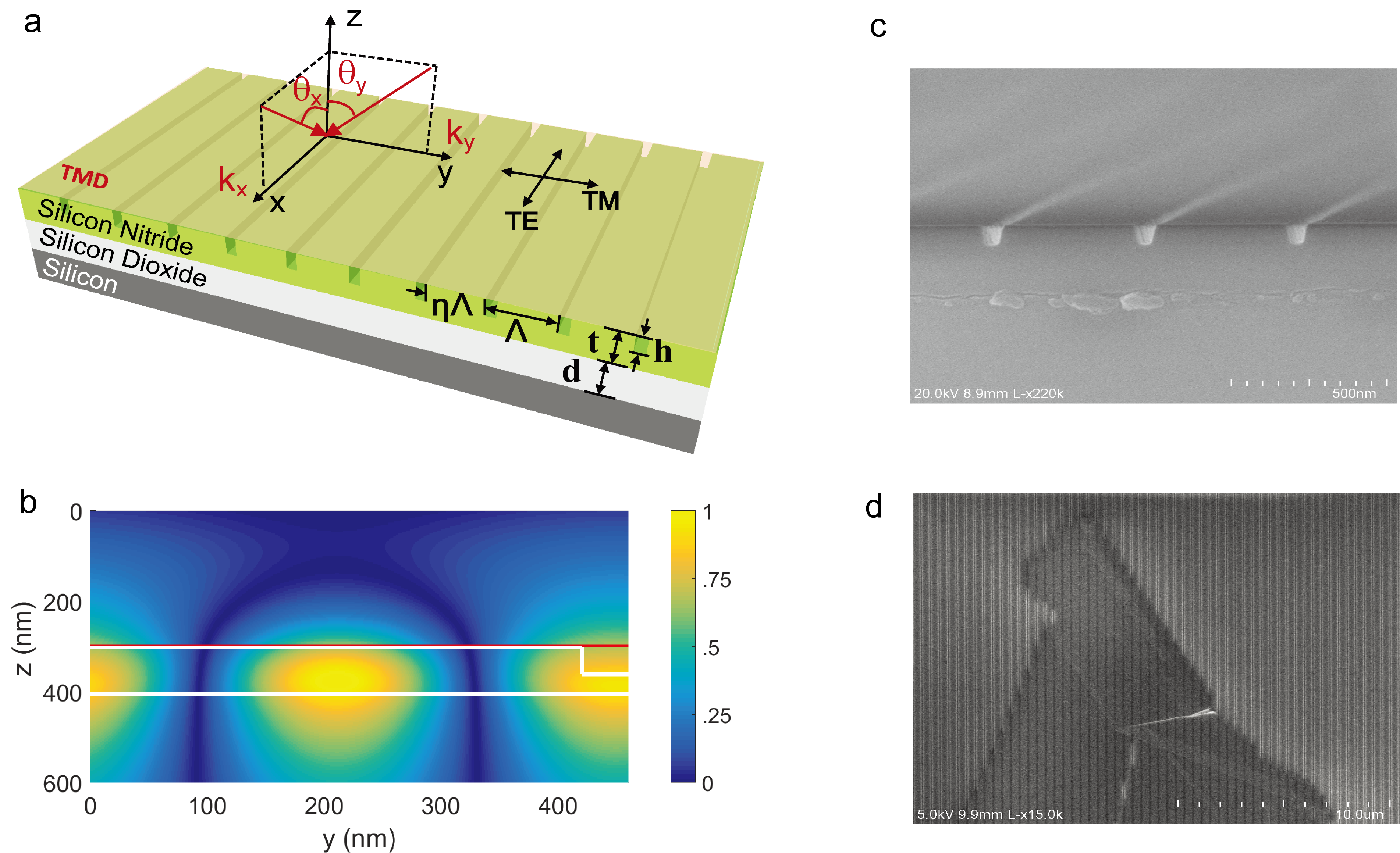}
	\caption{The vsWC-PC structure. (a) Schematic of the 1D PC with a monolayer TMD placed on top. The SiN PC has multiple design parameters, including the period $\Lambda$, filling factor $\eta$, total thickness $t$ and the grating thickness $h$. The SiO$_2$ capping layer has a thickness of $d$ .
(b) The TE-polarized electric field profile of the PC in the $y-z$ plane. The white lines mark the outline of the PC. The red line marks the position of the monolayer TMDs. (c) A side-view SEM image of the bare PC. (d) A top-view SEM image of the TMD laid on top of a PC. }
    \label{fig:sample}
\end{figure*}

We use two kinds of transition metal dichalcogenides (TMDs) as the active media: a monolayer of tungsten diselenide (WSe$_2$) or a monolayer of tungsten disulfide (WS$_2$). The monolayers are placed over a PC made of a silicon-nitride (SiN) grating, as illustrated in \ref{fig:sample}a. The total thickness of the grating $t$ is around 100~nm, much shorter than half a wavelength, making the structure an attractive candidate for compact, integrated polaritonics. In comparison, typical dielectric FP cavity structures are many tens of wavelengths in size. A schematic and scanning electron microscopy (SEM) images of the TMD-PC polariton device are shown in \cref{fig:sample}. More details of the structure and its fabrication are described in Methods. Since the grating is anisotropic in-plane, its modes are sensitive to both the propagation and polarization directions of the field. As illustrated in \cref{fig:sample}a, we define the direction along the grating bars as the $x$-direction, across the grating bars as the $y$-direction, and perpendicular to the grating plane as the $z$-direction. For the polarization, along the grating corresponds to transverse-electric (TE), and across the bar, transverse-magnetic (TM).  
%
The TM-polarized modes are far off resonance with the exciton. Hence TM excitons remain in the weak coupling regime, which provides a direct reference for the energies of the un-coupled exciton mode. We focus on the TE-polarized PC modes in the main text and discuss the TM measurements in the Supplementary Figure S1.

\subsection{WSe$_2$-PC polaritons}
\begin{figure*}[t]
	\includegraphics[width=\linewidth]{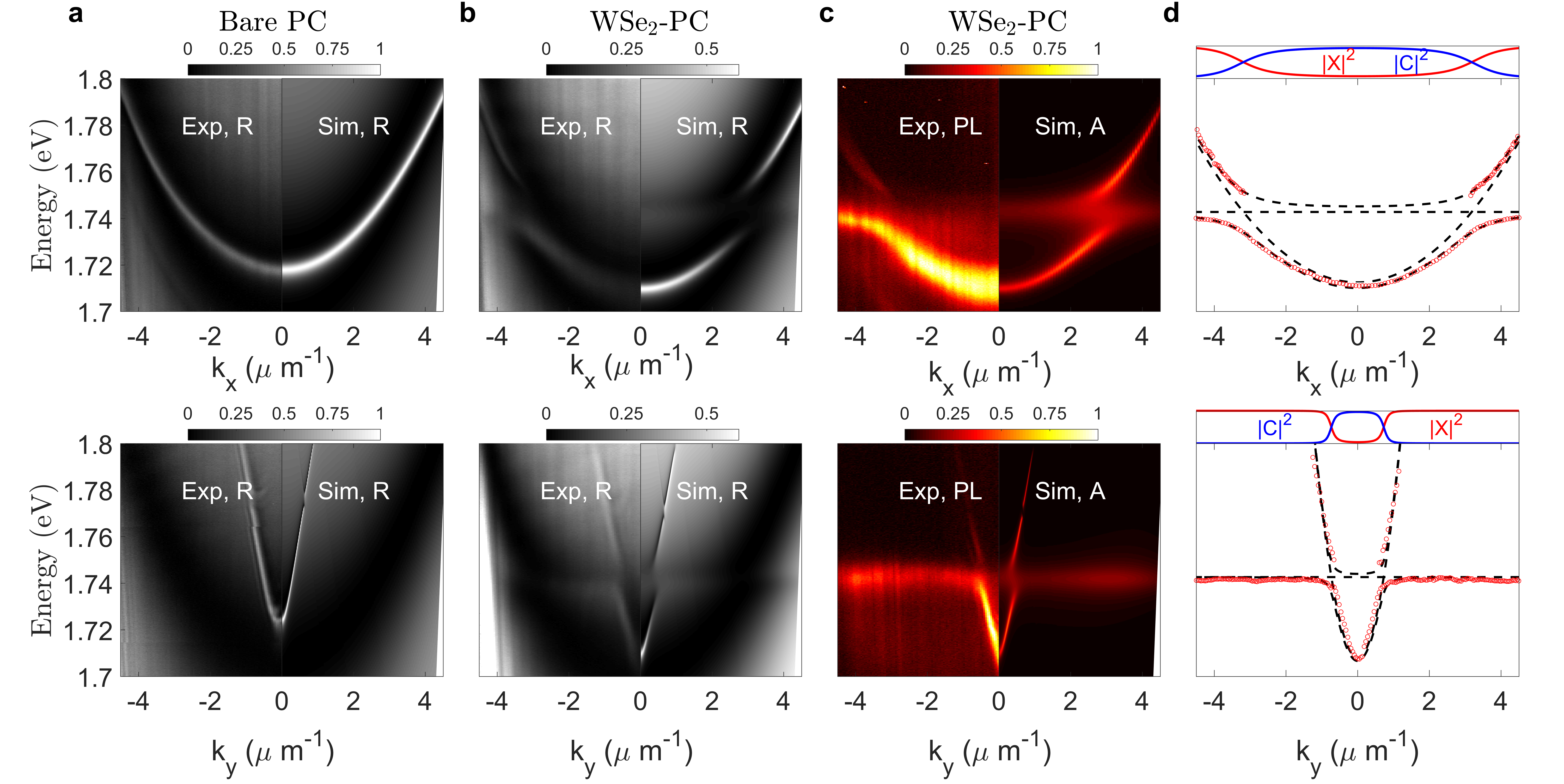}
	\caption{
		Strong coupling between TE-polarized WSe${_2}$ exciton and PC modes measured by angle-resolved reflectance and PL at 10~K. The top/bottom row shows the along-bar/cross-bar directions, respectively. The left/right panels of (a)-(c) show the measured/simulated results, respectively. (a) Angle-resolved reflectance spectra of the bare PC, showing a sharp, dispersive PC mode. (b) Angle-resolved reflectance spectra of the WSe${_2}$-PC integrated device, showing split, anti-crossing upper and lower polariton modes. (c) Angle-resolved PL data (left) compared with the simulated absorption spectra of the WSe${_2}$-PC integrated device, showing the same anti-crossing polariton modes as in (b). (d) The polariton energies $E_{LP,UP}$ vs. wavenumber $k_x$, $k_y$ obtained from the spectra in (c). The lines are fits to the LP and UP dispersion with the coupled harmonic oscillator model, giving a vacuum Rabi splitting of 18.4~meV and 16.1~meV for the along-bar (top) and cross-bar (bottom) directions, respectively. The corresponding Hopfield coefficients $|C^2|$ and $|X^2|$, representing the photon and exciton fractions in the LP modes, respectively, are shown in the top sub-plots. }
	\label{fig:WSe2}
\end{figure*}

We first characterize a monolayer WSe$_2$-PC device at 10~K.
The energy-momentum mode structures are measured via angle-resolved micro-reflectance (\cref{fig:WSe2}a-b)  and micro-PL (\cref{fig:WSe2}c) spectroscopy, in both the along-bar (top row) and across-bar (bottom row) directions. 
The data (left panels) are compared with numerical simulations (right panels), done with rigorous coupled wave analysis (RCWA).

Without the monolayer, a clear and sharp PC mode is measured with a highly anisotropic dispersion (\cref{fig:WSe2}a, left panels) and is well reproduced by simulation (\cref{fig:WSe2}a, right panels). The broad low-reflectance band in the background is an FP resonance formed by the SiO$_2$ capping layer and the substrate. The PC mode half linewidth is $\gamma_{cav}=6.5$~meV. This corresponds to a quality factor $Q$ or finesse of about 270, much higher than most TMD-cavities \cite{liu_strong_2014,sidler_fermi_2016,flatten_room-temperature_2016,lundt_room-temperature_2016,wang_coherent_2016,liu_strong_2016} and comparable to the best DBR-DBR ones \cite{dufferwiel_excitonpolaritons_2015,liu_control_2017}.

With a WSe$_2$ monolayer laid on top of the PC (\cref{fig:sample}c), two modes that anti-cross are clearly seen in both the reflectance and PL spectra (\cref{fig:WSe2}b-c) and match very well with simulations, suggesting strong coupling between WSe${_2}$ exciton and PC modes.
Strong anisotropy of the dispersion is evident comparing $E_{LP,UP}(k_x, k_y=0)$ (top row) and $E_{LP,UP}(k_x=0, k_y)$ (bottom row), resulting from  the anisotropic dispersion of the PC modes. 
Correspondingly, the effective mass and group velocity of the polaritons are also highly anisotropic, which provide new degrees of freedom to verify polariton condensation and to control its dynamics and transport properties \cite{dalfovo_theory_1999}.

To confirm strong-coupling, we fit the measured dispersion with that of coupled modes, and we compare the coupling strength and Rabi-splitting obtained from the fitting with the exciton and photon linewidth. In the strong coupling regime, the eigen-energies of the polariton modes $E_{LP,UP}$ at given in-plane wavenumber $k_{\parallel}$ and the corresponding vacuum Rabi splitting $2\hbar\Omega$ are given by:
\begin{align}
\label{eq:E_pol}
   E_{LP,UP} = & \frac{1}{2}\bigg[ E_{exc}+E_{cav}+i( \gamma_{cav}+\gamma_{exc} )/2 \bigg] \nonumber\\
    \pm&\sqrt{g^2+\frac{1}{4}\bigg[E_{exc}-E_{cav}+i(\gamma_{cav}-\gamma_{exc})\bigg]^2},
\end{align}
\begin{equation}
\label{eq:Rabi}
  2\hbar\Omega=2\sqrt{g^2-(\gamma_{cav}-\gamma_{exc}(T))^2/4}.
\end{equation}
Here $E_{exc}$ is the exciton energy, $\gamma_{exc}$ and $\gamma_{cav}$ are the half-widths of the un-coupled exciton and PC resonances, respectively, and $g$ is the exciton-photon coupling strength. A non-vanishing Rabi splitting $2\hbar\Omega$ requires  $g>|\gamma_{exc} - \gamma_{cav}|/2$; but this is insufficient for strong coupling. For the two resonances to be spectrally separable, the minimum mode-splitting needs to be greater than the sum of the half linewidths of the modes:
\begin{align}\label{eq:strongcoupling}
  2\hbar\Omega &> \gamma_{cav}+\gamma_{exc}, \textrm{or, } g >\sqrt{(\gamma_{exc}^2 + \gamma_{cav}^2)/2}.
\end{align}
In frequency domain, \cref{eq:strongcoupling} corresponds to requiring coherent, reversible energy transfer between the exciton and photon mode. 
%
%
We first fit our measured PL spectra to obtain the mode dispersion $E_{LP,UP}(k_{x,y})$, as shown by the symbols in \cref{fig:WSe2}d. We then fit $E_{LP,UP}(k_{x,y})$ with \eqref{eq:E_pol}, with $g$ and $E_{cav}(k_{x,y}=0)$ as the only fitting parameters. The exciton energy $E_{exc}$ and half-width $\gamma_{exc}$ are measured from the TM-polarized exciton PL from the same device, while the wavenumber dependence of $E_{cav}$ and $\gamma_{cav}$ are measured from the reflectance spectrum of the bare PC (Supplementary Figure S2b). We obtain $g=8.9\pm 0.23$~meV and $7.5\pm0.87$~meV for dispersions along $k_x$ and $k_y$, respectively, corresponding to a Rabi splitting of $2\hbar\Omega\sim$ 17.6~meV and 14.9~meV. In comparison, $\gamma_{exc}=5.7$~meV and $\gamma_{cav}=3.25$~meV. Therefore $g$ is much greater than not only $(\gamma_{exc}-\gamma_{cav)}/2=1.2$~meV  but also $\sqrt{(\gamma_{exc}^2+\gamma_{cav}^2)/2}=4.6$~meV, 
which confirms the system is well into the strong coupling regime. 

\subsection{Temperature dependence of WSe$_2$-PC polaritons}

At elevated temperatures, increased phonon scattering leads to faster exciton dephasing, which drives the system into the weak-coupling regime. We characterize this transition by the temperature dependence of the  WSe$_2$-PC system; we also show the effect of strong coupling on exciton quantum yield.

\begin{figure*}[t]
  \includegraphics[width=\linewidth]{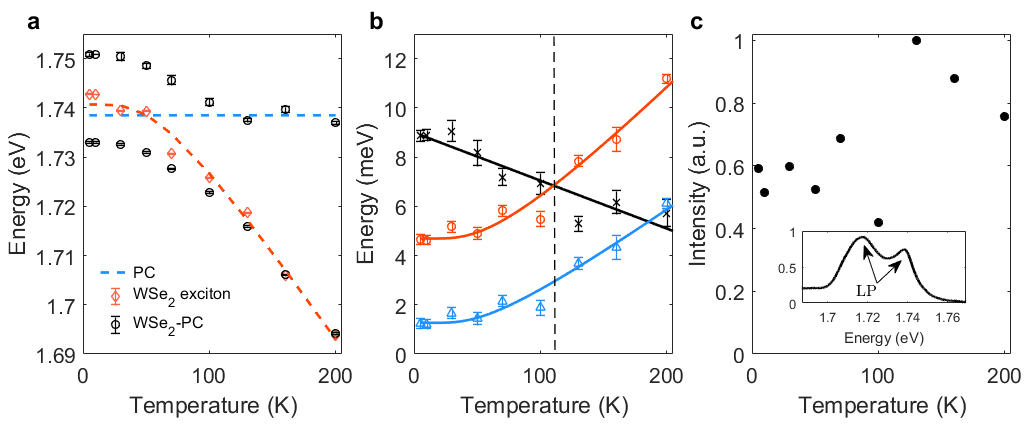}
	\caption{Temperature dependence of the WSe${_2}$-PC system.
(a) The temperature dependence of the exciton, cavity and polariton energies. The exciton energy $E_{exc}(T)$ is measured from the weakly coupled TM excitons (red circles) and fit by \cref{eq:E_T} (dashed blue line). The cavity energy $E_{cav}(T)$ is found to be approximately constant with temperature and indicated by the dashed blue line. The polariton energies $E_{LP,UP}(T)$ (black circles) are obtained from the PL spectra at $k_x=3.1\mu m ^{-1}, k_y=0\mu m ^{-1}$. They anti-cross and split from the exciton and PC mode energies at 100~K and below, showing strong coupling in this range.
(b) The strong coupling to weak coupling transition measured by the temperature dependence of $g$ (black stars), $\sqrt{(\gamma_{exc}^2+\gamma_{cav}^2)/2}$ (red circles), and $(\gamma_{exc}-\gamma_{cav})/2$ (blue triangles). The red and blue lines are fits by \cref{eq:dE_T} for $\gamma_{exc}$ while $\gamma_{cav}$ is approximately constant with temperature.  $g$ drops to below $\sqrt{(\gamma_{exc}^2+\gamma_{cav}^2)/2}$ at about 115~K, showing the transition to the weak coupling regime as indicated by the dashed line. Rabi-splitting persists up to about 185~K, till $g$ becomes smaller than  $(\gamma_{exc}-\gamma_{cav})/2$.
(c) Temperature dependence of the integrated PL intensity of the PC-WSe${_2}$ polariton, showing only mild change in intensity. The inset shows the spectrum at 10~K integrated over $\theta_x=-30^{\circ}$ to $ 30^{\circ} $ and over the spectral range shown. The higher energy side shoulder corresponds to exciton-like LP emission at large $\theta_x$ where the density of states is high.
}
	\label{fig:WSe2-T}
\end{figure*}

We measure independently the temperature dependence of the uncoupled excitons via TM exciton PL, the uncoupled PC modes via reflectance from the bare PC, and the coupled modes via PL from the WSe$2$-PC device. We show in \cref{fig:WSe2-T}a the results obtained for $k_x=3.1~\mu m ^{-1}, k_y=0~\mu m ^{-1}$ as an example.
For the uncoupled excitons, with increasing $T$, the resonance energy $E_{exc}(T)$ decreases due to bandgap reduction \cite{odonnell_temperature_1991}, as shown in \cref{fig:WSe2-T}a, while the linewidth $2\gamma_{exc}$ broadens due to phonon dephasing \cite{rudin_temperature-dependent_1990}, as shown in \cref{fig:WSe2-T}b. Both results are very well fitted by models for conventional semiconductors (see more details in Methods).
For the uncoupled PC modes, the energy $E_{cav}=1.74$~eV and half-linewidth $\gamma_{cav}=6.5$~meV change negligibly (Supplementary Figure S2c). The exciton and PC-photon resonances cross, as shown in \cref{fig:WSe2-T}a at around 50~K.
In contrast, the modes from the WSe$_2$-PC device anti cross between 10-100~K and clearly split from the uncoupled modes, suggesting strong-coupling up to 100~K. Above 130~K, it becomes difficult to distinguish the modes from WSe$_2$-PC device and the uncoupled exciton and photon modes, suggesting the transition to the weak-coupling regime.

We compare quantitatively in \cref{fig:WSe2-T}b the coupling strength $g$ with $\sqrt{(\gamma_{exc}(T)^2+\gamma_{cav}^2)/2}$ and  $(\gamma_{exc}-\gamma_{cav})/2$ to check the criterion given in ~\cref{eq:strongcoupling}. The strong coupling regime persists up to about 110~K, above which, due to the increase of the exciton linewidth, $g(T)$ drops to below $\sqrt{(\gamma_{exc}(T)^2+\gamma_{cav}^2)/2}$ and the system transitions to the weak-coupling regime, which corresponds well to the existence/disappearance of mode-splitting in \cref{fig:WSe2-T} below/above 110~K. On the other hand, $g>(\gamma_{exc}-\gamma_{cav})/2$ is maintained up to about 185~K. Between 110~K and 185~K, coherent polariton modes are no longer supported in the structure but mode-splitting remains in the reflectance spectrum (Supplementary Figure S3).

Importantly, the temperature dependence of the polariton PL intensity reveals that strong coupling enables significant enhancement of the quantum yield of WSe$_2$ at low temperatures. 
It has been shown that the quantum yield of the bright excitonic states are strongly suppressed by 10-100 fold in bare WSe$_2$ monolayers due to relaxation to dark excitons lying at lower energies than the bright excitons \cite{zhang_experimental_2015,wang_spin-orbit_2015}. In contrast, the WSe$_2$-PC polariton intensity decreases by less than two-fold from 200~K to 10~K. This is because coupling with the PC greatly enhances the radiative decay of the WSe$_2$ exciton-polariton states in comparison with scattering to the dark exciton states, effectively improving the quantum yield of the bright excitons.

\subsection{Room temperature WS$_2$-PC polaritons}
\begin{figure*}[t]
  \includegraphics[width=\linewidth]{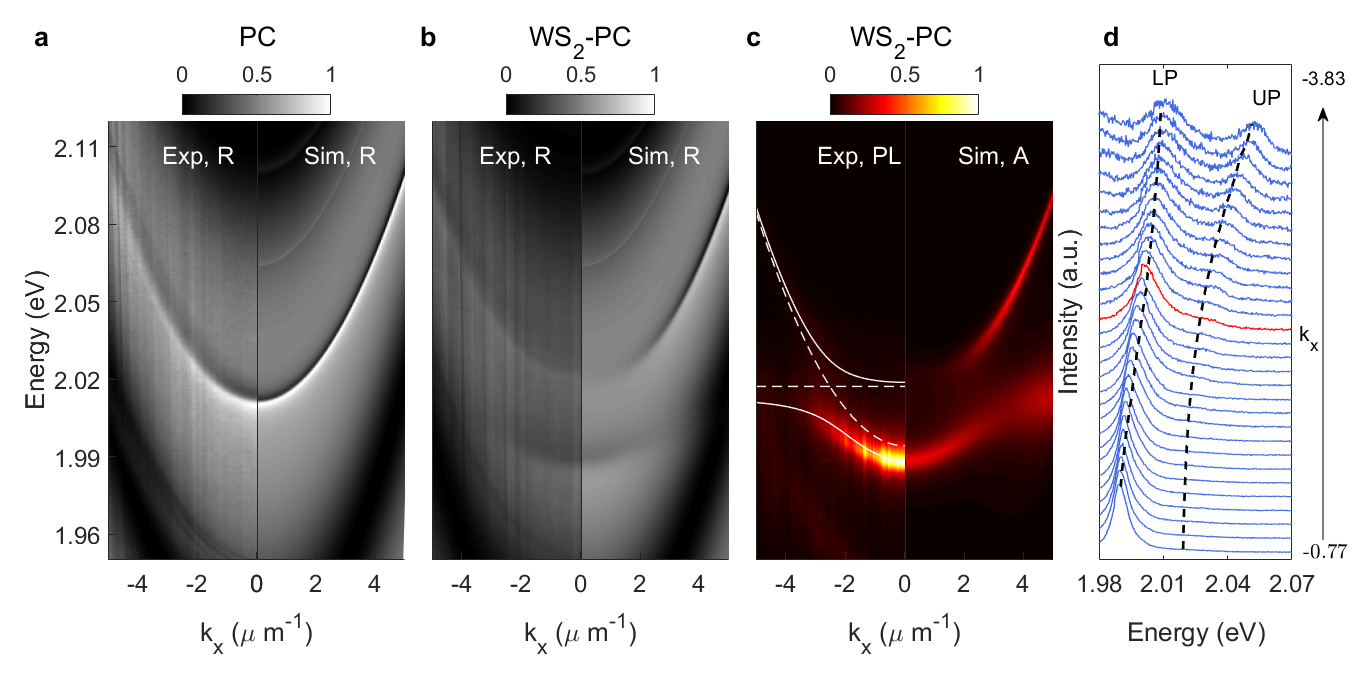}
	\caption{
Room-temperature strong coupling between TE-polarized WS${_2}$ exciton and PC modes measured by angle-resolved reflectance and PL. The left/right panels are the measured/simulated results, respectively.
(a) Angle-resolved reflectance spectra of the bare PC, showing a sharp, dispersive PC mode. (b) Angle-resolved reflectance spectra of the WS${_2}$-PC integrated device, showing split, anti-crossing upper and lower polariton modes. (c) Angle-resolved PL data (left) compared with the simulated absorption spectra of the WS${_2}$-PC integrated device, showing the same anti-crossing polariton modes as in (b). The solid lines are the fitted polariton dispersion, with a corresponding vacuum Rabi splitting of 22.2~meV. Dashed lines represent the exciton and cavity photon dispersion. (d) Normalized PL intensity spectra from (c) for $k_x=-0.77\mu m ^{-1}$ (bottom) to $k_x=-3.83\mu m ^{-1}$ (top). The red line marks the zero detuning at $-2.53\mu m ^{-1}$.
The dashed lines mark the fitted LP and UP positions, corresponding to the white solid lines marked in (c) .
}
    \label{fig:WS2}
\end{figure*}

To form exciton-polaritons at room temperature, we use WS${_2}$ because of the large oscillator-strength to linewidth ratio at 300~K compared to WSe$_2$ (Supplementary Figure S4). We use a 1D PC that matches the resonance of the WS$_2$ exciton at 300~K. The angle-resolved reflectance spectrum from the bare PC again shows a clear, sharp dispersion (\cref{fig:WS2}a). The broadband background pattern is due to the FP resonance of the substrate. With a monolayer of WS$_2$ placed on top, anti-crossing LP and UP branches form, as clearly seen in both the reflectance and PL spectra (\cref{fig:WS2}b-c). The data (left panels) are in excellent agreement with the simulated results (right panels). The dispersions measured from PL fit very well with the coupled oscillator model in \cref{eq:E_pol}, from which we obtain an exciton-photon interaction strength of $g=12.4\pm0.36$~meV, above $\gamma_{exc}=11$~meV, $\gamma_{cav}=4.5$~meV, and $\sqrt{(\gamma_{exc}^2+\gamma_{cav}^2)/2}=8.4 $~meV. The Rabi splitting is $2\hbar\Omega = 22.2$~meV. 

\subsection{Adjustable reflectance spectra with Fano resonances}
\begin{figure*}[t]
  \includegraphics[width=\linewidth]{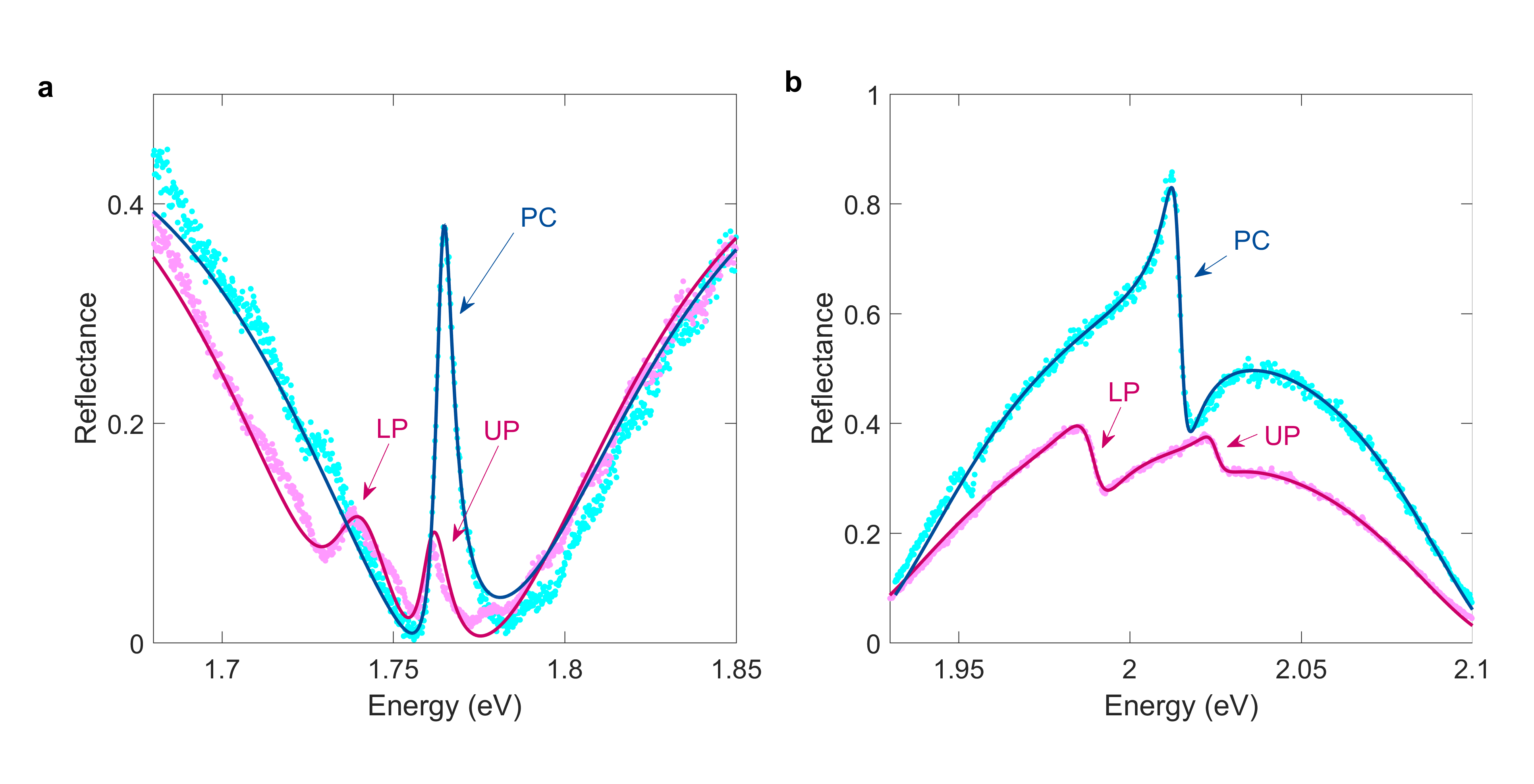}
	\caption{
Fano resonance in the reflectance of the (a) WSe${_2}$ and (b) WS${_2}$ systems, measured at $\theta_x=24^{\circ}$ and $6^{\circ}$, respectively. Blue dots are the reflectance spectra of the bare PCs, and red dots, the TMD-PC devices. The lines are comparison with Fano line shape given by \cref{eq:Fano}. The asymmetry parameters q for PC, LP, UP in WSe${_2}$ system are 5.0, 3.5, 4.1 respectively, and in WS${_2}$ system, are $1.16$, $0.92$, and $1.37$.
}
    \label{fig:fano}
\end{figure*}

Lastly, we look into two unconventional properties of the reflectance of the TMD-PC polariton systems: an adjustable reflectance background, and highly asymmetric Fano resonances. As shown in \cref{fig:WSe2} and \cref{fig:WS2}, a broadband background exists in the reflectance spectra for both WSe$_2$-PC and WS$_2$-PC polariton systems, arising from the FP resonances of the substrate. The height and width of this broadband background is readily adjusted by the thickness of the SiO$_2$ spacer layer, un-correlated with the quality factor of the PC modes or the lifetime of the polaritons. For example, the WSe$_2$-PC polaritons are in the low-reflectance region of the FP bands (\cref{fig:fano}a), while the WS$_2$-PC polaritons are in the high-reflectance region (\cref{fig:fano}b). In contrast, in conventional FP cavities, high cavity quality factor dictates that the polariton modes are inside a broad high-reflectance stop-band, making it difficult to excite or probe the polariton systems at wavelengths within the stop-band. The adjustability of the reflectance in PC-polariton systems will allow much more flexible access to the polariton modes and facilitate realization of polariton lasers, switches and other polariton nonlinear devices.

Another feature is the asymmetric Fano line shape of the PC and PC-polariton modes in the reflectance spectra (\cref{fig:fano}).
The Fano resonance arises from coupling between the sharp, discrete PC or PC-polariton modes and the continuum of free-space radiation modes intrinsic to the 2D-slab structure \cite{zhou_progress_2014}. Such Fano line shapes are readily tuned by varying the phase difference between the discrete mode and the continuum band. For example, the PC and WSe$_2$-PC polariton modes located at the valley of the FP band (\cref{fig:fano}a) have a nearly-symmetric Lorentzian-like line shape, but the PC and WS$_2$-PC polariton modes at the peak of the FP band feature a very sharp asymmetric Fano line shape (\cref{fig:fano}b). This is because of the $\pi$ phase difference between the peak and valley of the FP bands.

We compare the measured spectra with the standard Fano line shapes described by:
\begin{equation}
\label{eq:Fano}
  R=R_F\bigg(\frac{(\epsilon+q)^2}{\epsilon^2+1}-1\bigg)+R_{FP}+I_b,
\end{equation}
The first term describes the Fano resonance, where $R_{F}$ is the amplitude coefficient,  $q$ is the asymmetry factor, $\epsilon=\frac{\hbar(\omega-\omega_0)}{\gamma_0}$ is the reduced energy,  $\hbar\omega_0$ and $\gamma_0$ are the resonant energy and half linewidth of the discrete mode. $R_{FP}(\omega)$ and $I_b$ are the FP background reflectance and a constant ambient background, respectively. We use the transfer matrix method to calculate $R_{FP}$, then fit our data to \cref{eq:Fano} to determine the Fano parameters. For the WSe${_2}$-PC spectrum, we obtain $q_{cav}=5.0$, $q_{LP}=3.5$, $q_{UP}=4.1$, for the PC, LP and UP modes, respectively. The large values of $q$ suggest small degrees of asymmetry and line shapes close to Lorentzian, as seen in \cref{fig:fano}a. For the WS$_{2}$ device, we obtain $q_{cav}=1.16$, $q_{LP}=0.92$, and $q_{UP}=1.37$, which are close to 1. This corresponds to a much more asymmetric line shape with a sharp Fano-feature, as seen in \cref{fig:fano}b.
We note that, despite the striking Fano resonance in reflectance, strong coupling takes place only between the exciton and the sharp, discrete, tightly confined PC modes. This is evident from the symmetric line shape of the WS$_2$-polariton PL spectra \cref{fig:WS2}d. Fano resonances with polariton states as the discrete modes will enable control of the Fano line shape by angle and detuning. \cite{wang_exciton-polariton_2017}

\section{Discussion}
In short, we demonstrate integration of two of the most compact and versatile systems -- atomically thin vdWCs as the active media and PCs of deep sub-wavelength thicknesses as the optical structure -- to form an untra-compact and designable polariton system. TMD-PC polaritons were observed in monolayer WS$_2$ at room temperature and in WSe$_2$ up to 110~K, 
which are the highest temperatures reported for unambiguous determination of strong-coupling for each type of TMD, respectively. The TMD-PC polaritons feature highly anisotropic energy-momentum dispersions, adjustable reflectance with sharp Fano resonances, and strong suppression of non-radiative loss to dark excitons. These features will facilitate control and optimization of polariton dynamics for nonlinear polariton phenomena and applications, such as polariton amplifiers \cite{savvidis_angle-resonant_2000}, lasers \cite{deng_polariton_2003}, switches \cite{amo_exciton-polariton_2010} and sensors \cite{miroshnichenko_fano_2010,zhou_progress_2014}.

The demonstrated quasi-2D TMD-PC polariton system is readily extended to 0D, 1D and coupled arrays of polaritons \cite{zhang_zero-dimensional_2014,zhang_coupling_2015}. The 1D PC already has many design parameter for mode-engineering; 
it can be extended to 2D PCs for even greater flexibility, such as different polarization selectivity \cite{konishi_circularly_2011} for controlling the spin-valley degree of freedom \cite{xu_spin_2014}. The TMDs can be substituted by and integrated with other types of atomically-thin crystals, including black phosphorous for wide band-gap tunability\cite{li_direct_2016}, graphene for electrical control\cite{withers_light-emitting_2015}, and hexagonal boron-nitride for field enhancement.

PCs feature unmatched flexibility in optical-mode engineering, while vdWCs allow unprecedented flexibility in integration with other materials, structures, and electrical controls \cite{xia_two-dimensional_2014,mak_photonics_2016}. Combining the two in the strong coupling regime opens a door to novel polariton quantum many-body phenomenon and device applications \cite{karzig_topological_2015,bardyn_topological_2015,nalitov_spin-orbit_2015,liew_optical_2008,liew_single_2010,flayac_single_2016,joana_steady-state_2016,byrnes_mott_2010,douglas_quantum_2015,angelakis_quantum_2017}.

\section*{Methods}

\noindent\textbf{Sample fabrication.}
The devices shown in \cref{fig:sample} were made from a SiN layer grown by low pressure chemical vapor deposition on a SiO$_2$-capped Si substrate. The SiN layer was partially etched to form a 1D grating, which together with the remaining SiN slab support the desired PC modes. The grating was created via electron beam lithography followed by plasma dry etching.
Monolayer TMDs are prepared by mechanical exfoliation from bulk crystals from 2D Semiconductors and transferred to the grating using Polydimethylsiloxane (PDMS). For the WSe${_2}$ device, the grating parameters are: $\Lambda=468$~nm, $\eta=0.88$, $t=113$~nm, $h=60$~nm, $d= 1475$~nm. For the WS${_2}$ device, the grating parameters are:  $\Lambda=413$~nm, $\eta=0.83$, $t=78$~nm, $h=40$~nm, $d= 2000$~nm.

\noindent\textbf{Optical measurements.}
Reflection and PL measurements were carried out by real-space and Fourier-space imaging of the device. An objective lens with numerical aperture (N.A.) of 0.55 was used for both focusing and collection. For reflection, white light from a tungsten halogen lamp was focused on the sample to a beam size of $15~\mu$m in diameter. For PL, a HeNe laser (633~nm) and a continuous-wave solid state laser (532~nm) were used to excite the monolayer WSe${_2}$ and WS${_2}$, respectively, both with 1.5~mW and a $2~\mu$m focused beam size. The collected signals were polarization resolved by a linear polarizer then detected by a Princeton Instruments spectrometer with a cooled charge-coupled camera. 

\noindent\textbf{RCWA simulation.}
Simulations are carried out using an open-source implementation of RCWA developed by Pavel Kwiecien to calculate the electric-field distribution of PC modes, as well as the reflection and absorption spectra of the device as a function of momentum and energy. 
The indices of refraction of the SiO$_2$ and SiN are obtained from ellipsometry measurements to be $n_\text{SiO2} = 1.45+\frac{0.0053}{\lambda^2}$ and $n_{\text{SiN}}=2.0+\frac{0.013}{\lambda^2} $ , where $\lambda$ is the wavelength in the unit of $\mu$m. The WSe$_2$ and WS$_2$ monolayers were modelled with a thickness of .7 nm, and the in-plane permittivities were given by a Lorentz oscillator model:
$$\epsilon(E) = \epsilon_B + \frac{f}{E_x^2 - E^2 - i\Gamma E}.
$$
For WSe$_2$, we used oscillator strength $f_{\text{WS}_2}$=0.7~eV$^2$ to reproduce the Rabi splitting observed in experiments, exciton resonance $E_{\text{WSe}_2}=1.742$~eV and full linewidth $\Gamma_{\text{WSe}_2}=11.4$~meV based on TM exciton PL, and background permittivity $\epsilon_{B, \text{WSe}_2}=25$\cite{morozov_optical_2015}. Likewise, for WS$_2$, we used $f_{\text{WS}_2}$=1.85~eV$^2$, $E_{\text{WS}_2}=2.013$~eV and $\Gamma_{\text{WS}_2}=22$~meV measured from a bare monolayer, and $\epsilon_{B, \text{WS}_2}= 16$\cite{li_measurement_2014}.

\noindent\textbf{Modeling the temperature dependence of the WSe$_2$ exciton energy and linewidth.}\\
The exciton resonance energies redshift with increasing temperature as shown in \cref{fig:WSe2-T}a. It is described by the standard temperature dependence of semiconductor bandgaps\cite{odonnell_temperature_1991} as follows:
\begin{equation}
\label{eq:E_T}
E_{g}(T)=E_{g}(0)-S\hbar\omega\bigg[\coth\bigg(\frac{\hbar\omega}{2kT}-1\bigg)\bigg].
\end{equation}
Here E$_{g}(0)$ is the exciton resonance energy at $T=0$~K, $S$ is a dimensionless coupling constant, and $\hbar\omega$ is the average phonon energy, which is about 15~meV in monolayer TMDs\cite{ross_electrical_2013,arora_excitonic_2015}. The fitted parameters are: E$_{g}(0)$=1.741 and $S=2.2$, which agree with reported results\cite{ross_electrical_2013,arora_excitonic_2015}.

The exciton linewidth $\gamma_{exc}$ as a function of temperature can be described by the following model\cite{selig_excitonic_2016,arora_excitonic_2015}:
\begin{equation}
\label{eq:dE_T}
\gamma_{exc}=\gamma_{0}+c_{1}T+\frac{c_{2}}{e^{\hbar\omega/kT}-1}.
\end{equation}
Here $\gamma_{0}$ is the linewidth at 0 K, the term linear in T depicts the intravalley scattering by acoustic phonons, and the third term describes the intervalley scattering and relaxation to the dark state through optical and acoustic phonons\cite{selig_excitonic_2016}.The average phonon energy is $\hbar\omega$=15~meV. The fitted parameters are: $\gamma_{0}=11.6$~meV, $c_{2}=25.52$~meV, and $c_{1}$ is negligibly small in our case\cite{arora_excitonic_2015}.

\section*{Acknowledgment}
LZ, RG and HD acknowledge the support by the Army Research Office under Awards W911NF-17-1-0312 and the Air Force Office of Scientific Research under Awards FA9550-15-1-0240. WB and ET acknowledge the support by National Science Foundation  Grant EECS-1610008. The fabrication of the PC was performed in the Lurie Nanofabrication Facility (LNF) at Michigan, which is part of the NSF NNIN network.

\bibliographystyle{nature3}
\bibliography{Reference}

\end{document}